\documentclass[reprint, amsmath,amssymb, aps, prl, shortbibliography, lengthcheck]{revtex4-1}

\pdfoutput=1
\usepackage{graphicx}
\usepackage{dcolumn}
\usepackage{bm}
\usepackage{epsfig}

\begin{document}

\title{Millikelvin Reactive Collisions between Sympathetically-Cooled Molecular Ions and Laser-Cooled Atoms in an Ion-Atom Hybrid Trap}

\author{Felix H.J. Hall}
\affiliation{Department of Chemistry, University of Basel, Klingelbergstrasse 80, 4056 Basel, Switzerland}
\author{Stefan Willitsch}
\email{stefan.willitsch@unibas.ch}
\affiliation{Department of Chemistry, University of Basel, Klingelbergstrasse 80, 4056 Basel, Switzerland}

\date{\today}

\begin{abstract}
We report on a study of cold reactive collisions between sympathetically-cooled molecular ions and laser-cooled atoms in an ion-atom hybrid trap. Chemical reactions were studied at average collision energies $\langle E_\text{coll}\rangle/k_\text{B}\gtrsim20$~mK, about two orders of magnitude lower than has been achieved in previous experiments with molecular ions. Choosing N$_2^+$+Rb as a prototypical system, we find that the reaction rate is independent of the collision energy, but strongly dependent on the internal state of Rb. Highly efficient charge exchange about four times faster than the Langevin rate was observed with Rb in the excited $(5p)~^2P_{3/2}$ state. This observation is rationalized in terms of a capture process dominated by the charge-quadrupole interaction and a near resonance between the entrance and exit channels of the reaction. Our results provide a test of classical models for reactions of molecular ions at the lowest energies reached thus far.
\end{abstract}

\maketitle

The combination of radiofrequency (RF) ion traps with magneto-optical or optical-dipole traps for the simultaneous confinement of cold ions and atoms has recently enabled the study of ion-neutral interactions at temperatures as low as a few millikelvin \cite{grier09a, zipkes10a, schmid10a, hall11a, rellergert11a, ravi12a, ratschbacher12a}. Hybrid trapping techniques have opened up perspectives for the exploration of new mesoscopic quantum systems \cite{cote02a}, for quantum interfaces between atoms and ions \cite{idziaszek07a} and for new methods to cool ions to ultralow temperatures \cite{hudson09a, ravi12a}. Moreover, chemical experiments with ions \cite{zipkes10b, hall11a, rellergert11a, ratschbacher12a} have started to approach an energy regime so far restricted to neutrals \cite{weiner99a, ospelkaus10b, ni10a, henson12a} in which the quantum character of the collision can strongly influence the chemical dynamics \cite{belyaev12a, gao10a, gao11a}.

Until now, studies of reactive processes in hybrid traps have been restricted to atomic ions and neutral atoms. Despite the apparent simplicity of these collision systems, a diverse range of chemical phenomena has been observed. The effects range from the prominent role of light-assisted processes in the radiative formation of molecules and the promotion of charge exchange in Ca$^+$+Rb \cite{hall11a}, the occurrence of unusually fast reaction rates in Yb$^+$+Ca \cite{rellergert11a} and the influence of the ion's hyperfine state on the reactivity in Yb$^+$+Rb \cite{ratschbacher12a}. However, experiments with atomic species can only cover a fraction of the wealth of chemical phenomena, and it is necessary to extend studies to molecules to be able to fully explore the diversity of reactive effects at ultralow energies.

In the present work, we extend ion-neutral hybrid-trapping techniques to enable the study of cold reactive collisions with molecular species by sympathetically cooling molecular ions via their interaction with laser-cooled atomic ions \cite{willitsch12a} and immersing them in a cloud of ultracold atoms confined in a magneto-optical trap (MOT). To our knowledge, this is the first time that chemical reactions with molecular ions have been studied at collision energies down to $\langle E_\text{col}\rangle/k_\text{B}\approx20$~mK, about two orders of magnitude lower than has been achieved in previous cold-collision experiments \cite{willitsch08a, smith98a, otto08a}. By choosing sympathetically-cooled N$_2^+$ ions and laser-cooled Rb atoms as a prototypical system, we characterize important features of cold reactive collisions between molecular ions and alkali atoms. We observe charge exchange between the two species and a strong dependence of the reaction rate on the electronic state of Rb. Whereas reactions with Rb atoms in the $(5p)~^2P_{3/2}$ state occur fast consistent with a classical capture model, reactions with ground-state Rb atoms were found to be at least two orders of magnitude slower. The fast excited-state rate is explained by a long-range capture of the atom by the molecular ion dominated by the long-range charge-quadrupole interaction and a near resonance between the electronic entrance and exit channels of the system at short range. Our results highlight the importance of resonant effects in cold ion-neutral collisions and provide a test of classical capture models for reactions of molecular ions at the lowest energies reached thus far.

\begin{figure}[t]
\epsfig{file=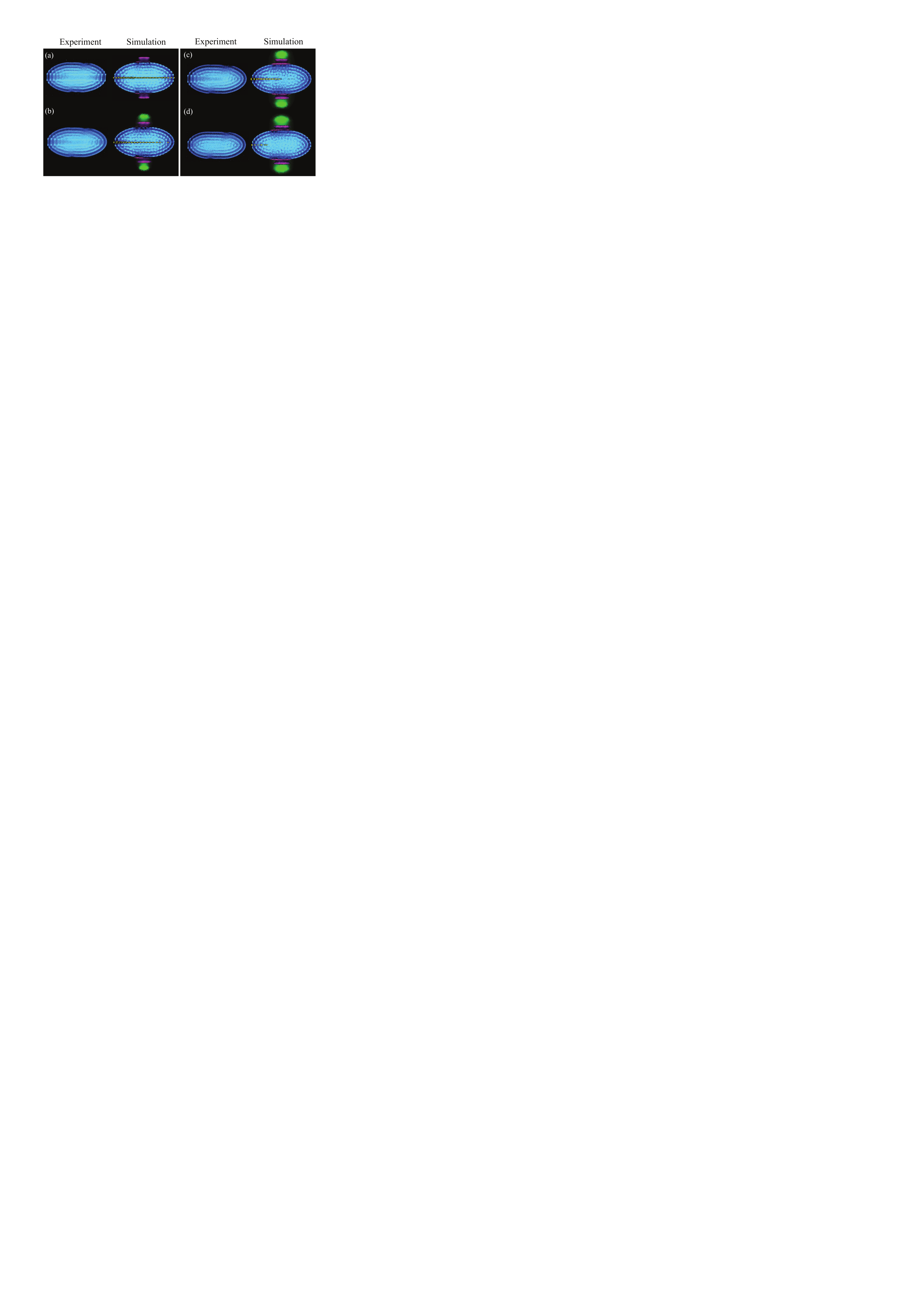,width=\columnwidth}
\caption{\label{rxn} Experimental false-color laser-cooling-fluorescence images of Coulomb crystals recorded over the course of reaction with ultracold Rb atoms. Sympathetically-cooled ions are made visible in molecular dynamics simulations. Color code: blue: laser-cooled Ca$^+$ ions; yellow: N$_2^+$ reactant ions; green: Rb$^+$ product ions; magenta: CaH$^+$, CaOH$^+$. See text for details.}
\end{figure}

Our hybrid-trapping apparatus has already been described in Ref. \cite{hall11a}. In the present study, N$_2^+$ ions were generated in a linear radiofrequency (RF) ion trap from room-temperature background nitrogen gas using [2+1] resonance-enhanced multiphoton-ionization (REMPI) via the $Q$ branch of the vibrationless $X~^1\Sigma_g^+\rightarrow a^{\prime\prime}~^1\Sigma^+_g$ transition \cite{tong11a}. The ions were subsequently thermalized through collisions with the background gas \cite{tong11a}.

After their generation, the N$_2^+$ ions were sympathetically cooled by the interaction with laser-cooled Ca$^+$ ions to form Coulomb crystals \cite{willitsch12a}. Images of the crystals were obtained by collecting the spatially resolved laser-cooling fluorescence of the Ca$^+$ ions using a CCD camera coupled to a microscope. The number of laser- and sympathetically-cooled ions in the crystals were determined by comparisons of the experimental images with molecular-dynamics (MD) simulations as described in Refs. \cite{bell09a, tong10a}. 

After ion loading, the Coulomb crystals were translated along the axis of the linear RF trap using a static electric field to be overlapped with the ultracold atoms. Clouds of ultracold Rb atoms at densities $n_\text{Rb}= 4\times 10^{8}$~cm$^{-3}$ were generated by laser-cooling in the MOT overlapped with the ion trap \cite{hall11a}. Rate constants for reactive collisions between ions and atoms were determined by monitoring the decrease of the number $N_{\text{N}_2^+}$ of N$_2^+$ reactant ions as a function of the reaction time $t$ and fitting the data to an integrated pseudo-first-order rate law $\ln[N_{\text{N}_2^+}(t)/N_{\text{N}_2^+}(t\text{=}0)]=-k_\text{pfo}t$. Second-order rate constants $k$ were determined from the pseudo-first-order constants $k_\text{pfo}$ using $k=k_\text{pfo}/n_\text{Rb}$. Both the sympathetically-cooled N$_2^+$ and the laser-cooled Ca$^+$ ions react with Rb. However, the total rate for Ca$^+$+Rb is two orders of magnitude smaller \cite{hall11a} than for N$_2^+$+Rb so that the reorganization of the Coulomb crystals caused by loss of Ca$^+$ could be neglected over the timescale of the rate measurements for N$_2^+$.

Fig. \ref{rxn} (a)-(d) shows a sequence of false-color Coulomb crystal images and their MD simulations recorded over the course of a typical reaction experiment. The laser-cooling fluorescence of the Rb atoms was blocked by a color filter so that only the Ca$^+$ fluorescence (blue) is visible in the images. The N$_2^+$ ions can only indirectly be seen as a dark non-fluorescing region in the center of the crystals, but have been made visible in yellow in the simulations. Initially, Ca$^+$ Coulomb crystals containing $\approx$1000~ions were loaded with typically $\approx$25 sympathetically-cooled N$_2^+$ ions (Fig. \ref{rxn} (a)). Over the course of the experiment, the N$_2^+$ ions were removed from the crystals by charge-exchange collisions with Rb atoms resulting in a reduction of the dark crystal core (Figs. \ref{rxn} (b)-(d)). The Rb$^+$ product ions (green in the simulations) remained trapped and were sympathetically cooled by the Ca$^+$ ions. The magenta-colored ions in the simulated images represent CaH$^+$ and CaOH$^+$ which are impurities formed by reactions of Ca$^+$ with background H$_2$ and H$_2$O gas, respectively.

\begin{figure}[t]
\epsfig{file=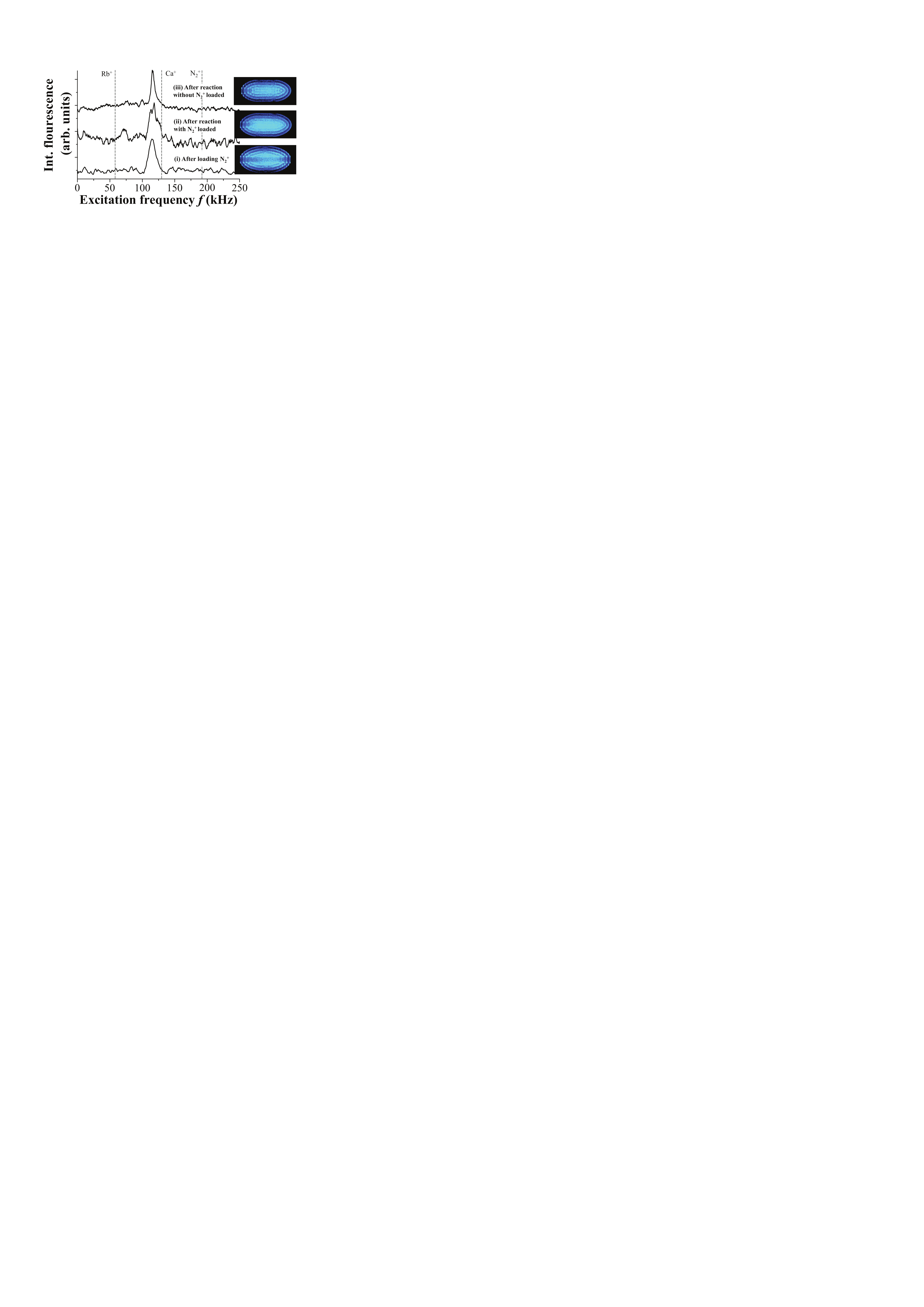,width=\columnwidth}
\caption{\label{ms} Resonant-excitation mass spectra of Coulomb crystals (insets) loaded with sympathetically-cooled N$_2^+$ ions (i) before reaction with ultracold Rb atoms and (ii) after reaction of the N$_2^+$ ions with Rb. The dashed vertical lines indicate the single-ion resonance frequencies of the relevant ion species. The weak feature around 70 kHz in (ii) is identified with the Rb$^+$ product ions generated by charge exchange of N$_2^+$ with Rb. (iii) Spectrum of a crystal prepared without N$_2^+$ ions recorded after the same time of reaction with Rb as in (ii) illustrating the effect of the slow background reaction Ca$^+$+Rb$\rightarrow$Ca+Rb$^+$. See text for explanation.}
\end{figure}

To prove the chemical identity of the reaction products, resonant-excitation mass spectra of the Coulomb crystals were recorded \cite{roth07a}. Briefly, an additional RF drive voltage applied to the trap electrodes was used to resonantly excite the radial motion of specific ion species. The motion of the excited ions couples to the laser-cooled Ca$^+$ ions via the Coulomb interaction and leads to a modulation of their laser-cooling fluorescence around the relevant resonant excitation frequencies. Fig. \ref{ms} shows mass spectra of representative Coulomb crystals (i) before and (ii) after reaction. Additionally, a pure Ca$^+$ crystal  (iii) was reacted for the same duration as in (ii) to illustrate the effect of the slow background reaction Ca$^+$ + Rb. The insets show images of the corresponding crystals and the vertical lines indicate the theoretical single-ion resonance frequencies of the relevant ion species. 

Whereas the spectrum recorded before reaction only shows a single resonance corresponding to the excitation of the Ca$^+$ ions, the spectrum after reaction in (ii) shows an additional weak feature around 70~kHz assigned to the sympathetically-cooled Rb$^+$ product ions. Note that the positions of the resonances in a Coulomb crystal are slightly shifted from the single-ion frequencies because of Coulomb-coupling effects \cite{roth07a}. Resonances for the CaOH$^+$, CaH$^+$ and the reactant N$_2^+$ ions, as well as for the Rb$^+$ ions formed by the background reaction Ca$^+$ + Rb in (iii) could not be observed in the spectra under the present conditions. MD simulations revealed that their numbers were too small to exert any noticeable back action on the Ca$^+$ fluorescence at the excitation amplitudes used. Higher amplitudes would have led to the melting of the crystals around the position of the Ca$^+$ resonance and were therefore precluded. 

\begin{figure}[t]
\epsfig{file=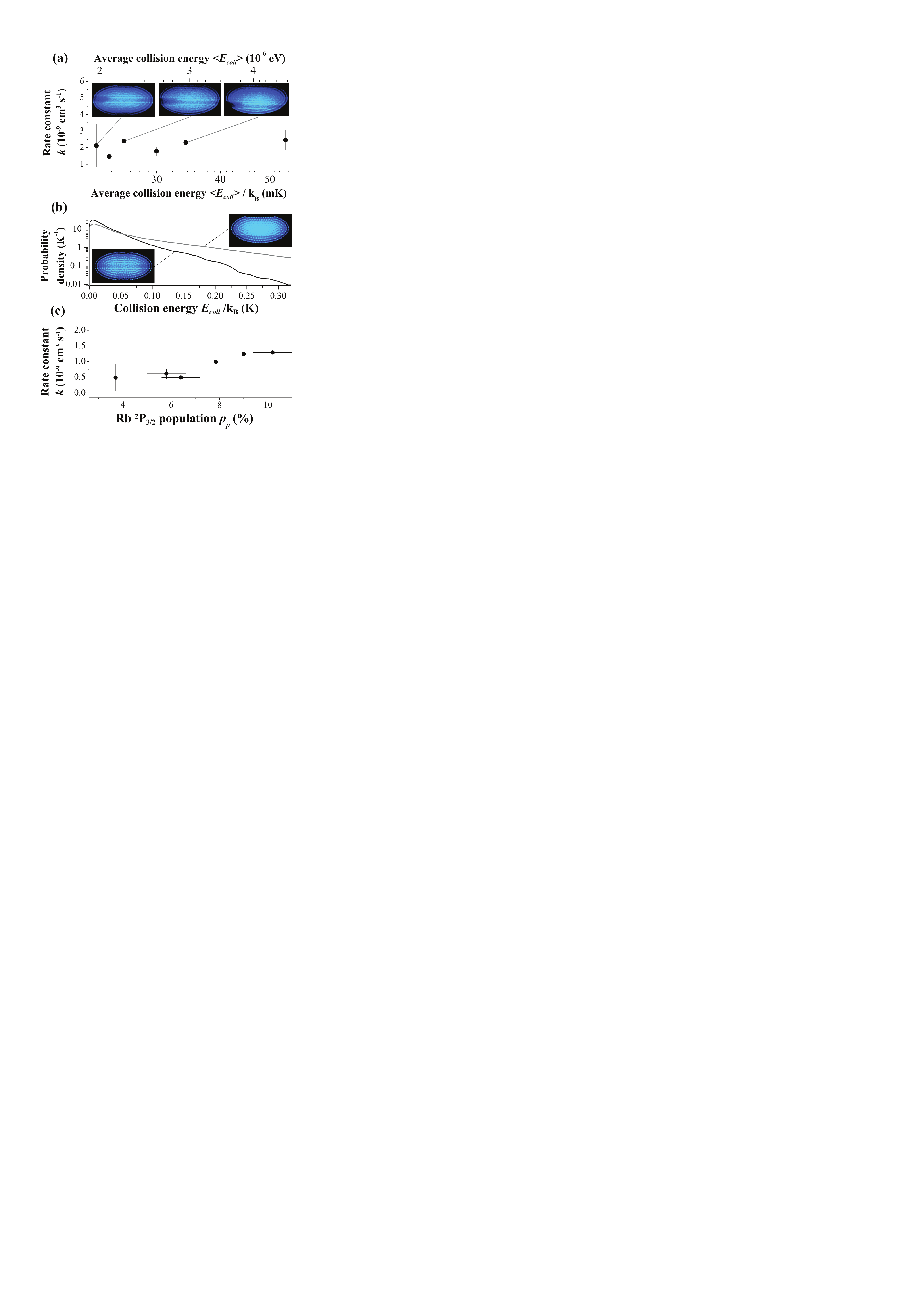,width=0.9\columnwidth}
\caption{\label{rates} (a) Rate constants for N$_2^+$+Rb as a function of the collision energy. Each data point represents an average over three independent measurements, the error bars indicate the statistical uncertainty $(2\sigma)$. The insets show the images of the Coulomb crystals corresponding to the three data points indicated. (b) N$_2^+$ ion kinetic-energy distributions from MD simulations corresponding to the data points with the lowest (black) and highest energies (grey) displayed in (a). The insets show the relevant simulated crystal images. (c) Rate constant as a function of the population in the Rb $(5p)~^2P_{3/2}$ state.}
\end{figure}

Fig. \ref{rates} (a) shows the dependence of the rate constant on the collision energy. In the experiments, the Rb kinetic energies ($\langle E_\text{kin}\rangle/k_\text{B}\approx 200\mu$K) were much smaller than the ion energies ($\gtrsim20$~mK) so that the collision energies were entirely dominated by the contribution of the ions. The motion of the ions can be separated into two different components, a slow secular (thermal) motion in the time-averaged pseudopotential of the trap and a fast micromotion driven by the RF fields \cite{willitsch12a}. The secular energies ($\langle E_\text{sec}\rangle /k_\text{B}$=12~mK for Ca$^+$ and 14~mK for N$_2^+$) and micromotion energies were determined from the ion trajectories obtained in the MD simulations of the crystal images \cite{bell09a}. The effective (total) kinetic energy $E_\text{tot}$ of the ions was governed by their micromotion energy which vanishes for ions located on the central trap axis and increases quadratically with the distance \cite{willitsch12a}. The lowest achievable energies are limited by the precision with which the N$_2^+$ ions can be localized on the axis. The axialization procedures adopted in the present experiments (to be described in detail in Ref. \cite{hall12a}) allowed for a micromotion energy limit corresponding to 17~mK.

For the measurements reported in Fig. \ref{rates} (a), Coulomb crystals with strings of nitrogen ions (like the ones shown in Fig. \ref{rxn} (a)) were aligned parallel to the trap axis so that all ions exhibit approximately the same micromotion energy \cite{bell09a}. The collision energies were varied by displacing the Coulomb crystals from the axis using static electric fields causing the N$_2^+$ ions to move into regions where they acquire an increased micromotion energy (see insets). The collision-energy distributions obtained from the MD simulations which correspond to the data points with the lowest and highest energies in Fig. \ref{rates} (a) are displayed in Fig. \ref{rates} (b) for reference along with the corresponding simulated images. As can be seen from Fig. \ref{rates} (a),  within the uncertainty limits the rate constant is essentially independent of the collision energy in the interval studied.

Because the ultracold atoms are constantly excited during laser cooling, a fraction of the reactions occurs with Rb atoms in the $(5p)~^2P_{3/2}$ state. The population in the excited state was adjusted by varying the intensity of the Rb cooling laser \cite{hall11a}. The  rate constant shows an approximately linear increase with the $(5p)~^2P_{3/2}$ population (Fig. \ref{rates} (c)). In analogy to Ref. \cite{hall11a}, a kinetic model of the form $k=\tfrac{1}{2}\big[k_s(1+p_s)+k_pp_p \big]$, where $k_{s,p}$ and $p_{s,p}$ denote the rate constants and populations in the Rb $(5s,5p)$ states, respectively, was fitted to the data. The fit yielded $k_p=2.4(13)\times10^{-8}$~cm$^3$s$^{-1}$ and an upper bound $k_s\leq 2\times10^{-10}$~cm$^3$s$^{-1}$. The uncertainty of $k_p$ is dominated by a systematic error of $\approx 50\%$ in the Rb density.

\begin{figure}[t]
\epsfig{file=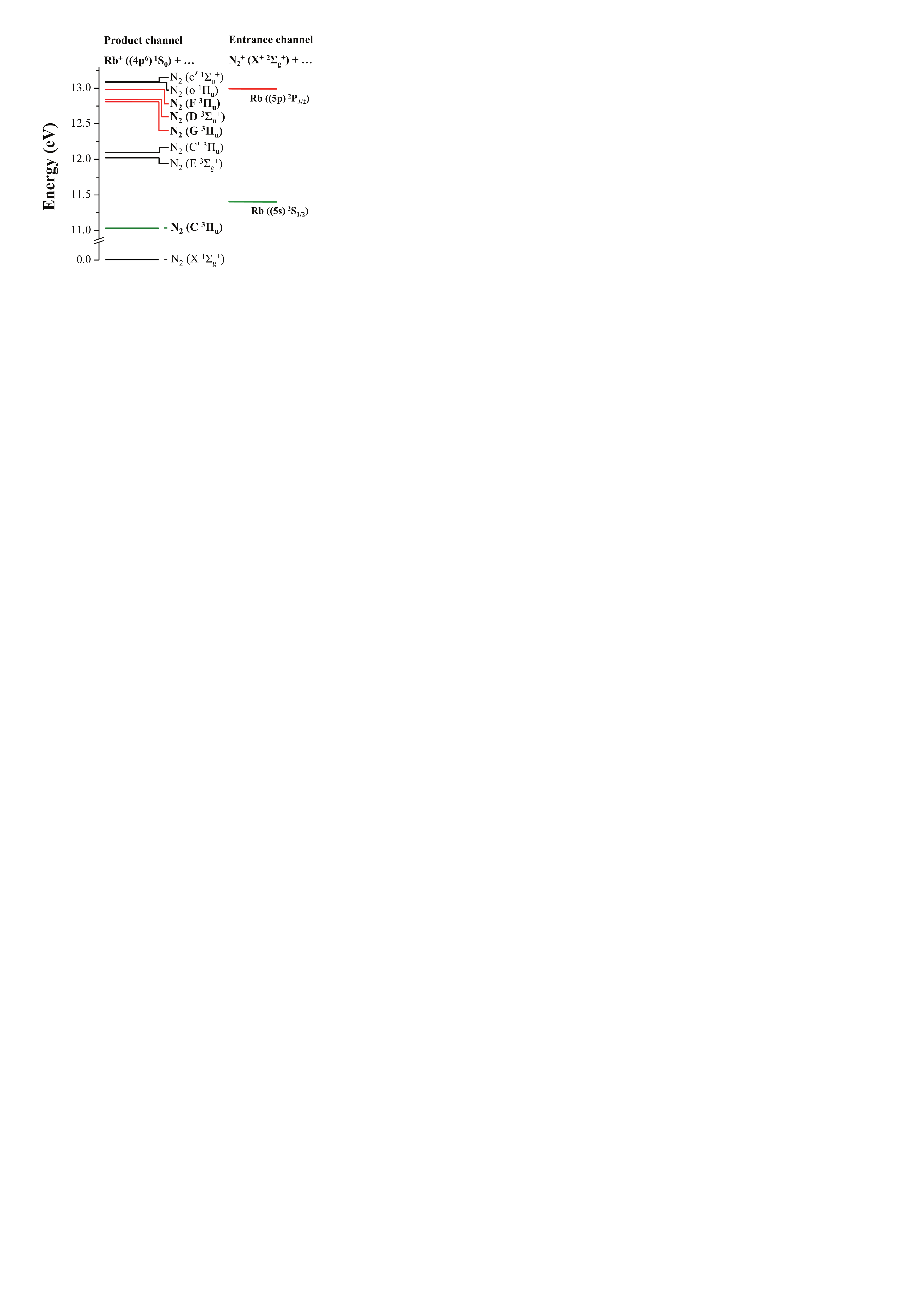,width=0.8\columnwidth}
\caption{\label{channels} Energy diagram of relevant electronic entrance and product channels for N$_2^+$+Rb. The closest product channels associated with the N$_2^+$+Rb$(5s)$ and $(5p)$ entrance channels are indicated in green and red, respectively.}
\end{figure}

Because of the large difference between the ionization energies of N$_2$ and Rb, a minimum of 11.4~eV of energy are released in the reaction. By comparison, the rotational-vibrational energy of the N$_2^+$ ions ($\approx0.025$~eV for a room-temperature distribution) as well as the collision energy ($\approx10^{-6}$~eV) are small, so that the available energy for the reaction is entirely dominated by the electronic contribution. Fig. \ref{channels} shows an energy diagram depicting the relevant electronic entrance and product channels (see, e.g., Ref. \cite{lofthus77a} for a comprehensive overview of the electronic states of N$_2$). 

Previous studies using keV ion beams showed that charge exchange between alkali atoms and N$_2^+$ is most efficient in a near-resonant transfer of the electron into a Rydberg state of the neutral product molecule which is built on an ion core with the same electronic configuration as the reactant ion \cite{vanderkamp94a}. In the present case, the N$_2^+(X^+)$+Rb$(5s)$ entrance channel is closest in energy to a product channel forming N$_2$ in the $C~^3\Pi_u$ electronic state (energy mismatch $\Delta E=372$~meV, see Fig. \ref{channels}). However, electron capture by N$_2^+(X^+)$ (molecular orbital configuration $(2\sigma_g)^2(2\sigma_u)^2(1\pi_u)^4(3\sigma_g)^1$) to form N$_2$ in the $C$ state ($(2\sigma_g)^2(2\sigma_u)^1(1\pi_u)^4(3\sigma_g)^2(1\pi_g)^1$) entails a significant re-arrangement of the electronic configuration of the molecular core. This process is therefore expected to be inefficient, in line with the present observation of a slow reaction rate in this channel. On the other hand, the excited N$_2^+(X^+)$+Rb$(5p)$ entrance channel is near resonant with product channels forming N$_2$ in the the close-lying $G~^3\Pi_u, D~^3\Sigma_u^+$ and $F~^3\Pi_u$ states (energy mismatches $\Delta E$=183~meV, 151~meV and 8~meV). These heavily mixed states \cite{lewis08a} are among the lowest Rydberg states of N$_2$ built on the N$_2^+~X^+~^2\Sigma^+_g$ and $A^+~^2\Pi_u$ ion cores so that capture of the electron into these states is expected to be highly efficient, in agreement with the fast rate observed in the excited channel. The Rydberg molecules subsequently predissociate into the atomic fragments N$(^4S)$+N$(^2D)$ \cite{vanderkamp94a} which, however, cannot be detected in the present experiment. 

In previous studies on cold ion-neutral collisions in hybrid traps \cite{grier09a, ratschbacher12a}, the rates were found to be in agreement with a limiting value set by Langevin theory \cite{gioumousis58a}. The rate constant for the excited channel determined above ($k_p=2.4(13)\times10^{-8}$~cm$^3$s$^{-1}$), however, is about four times larger than the Langevin value ($k^{(L)}=6.6\times10^{-9}$~cm$^3$s$^{-1}$). This finding suggests that intermolecular forces other than the Langevin (charge-induced dipole) interaction play a role. In the present case, the most likely candidate is the interaction of the charge of the ion with the permanent electric quadrupole moment of Rb generated by the anisotropic charge distribution in the $(5p)~^2P_{3/2}$ state \cite{pirani06a}.

The rate constant was calculated with a classical model based on an interaction potential \cite{krych11a} incorporating the Langevin and charge-quadrupole interactions. The quadrupole moment of Rb~$(5p)~^2P_{3/2}$ was computed within the single-electron approximation following Ref. \cite{sobelman79a}. In line with the reaction mechanism discussed above, a capture approximation \cite{dashevskaya03a} was adopted, i.e., the reaction was assumed to be governed by the long-range interactions and to occur with unit probability once the reaction complex was formed. Within this framework, the classical reaction cross section $\sigma^{(c)}=\pi b^2$ was calculated from the maximum possible impact parameter $b$ for which the height of the barrier in the centrifugally corrected potential does not exceed the collision energy \cite{levine05a}. From $\sigma^{(c)}$, the corresponding rate constant at a collision energy $E_\text{coll}/k_\text{B}=23$~mK was obtained to be $k_p^{(c)}=1.7\times10^{-8}$~cm$^3$s$^{-1}$, which agrees with the experimental value within its uncertainty limits. 

Notwithstanding the approximations inherent in the model (capture approximation, neglect of quantum effects and the molecular structure), the result underlines the necessity to reach beyond the Langevin picture to explain the dynamics in the present system. It also suggests that - at least for the present case - classical capture models seem to adequately reproduce the kinetics in the millikelvin regime. This conclusion is consistent with the theoretical results of Refs. \cite{dashevskaya03a, gao11a} which predict a robust performance of classical capture models for ion-neutral reactions down to ultracold ($<$mK) temperatures for all but the lightest systems. This behavior was rationalized in Ref. \cite{dashevskaya03a} in terms of a cancellation of tunneling and reflection effects at the centrifugal barrier.  

In conclusion, we have studied reactions between sympathetically-cooled N$_2^+$ ions and ultracold Rb atoms at average collision energies down to $\approx$20~mK in an ion-atom hybrid trap. Our results highlight the importance of resonant electronic effects in electron-transfer processes between ultracold alkali atoms and cold molecular ions. A classical capture model was found to adequately describe the kinetics down to the lowest collision energies for channels in which the short-range electron-transfer probability is expected to be close to unity. Whether short-range effects such as the rotational-vibrational state of the molecular ion which are not accounted for in the capture picture play a role in the present system remains to be established in further studies performed with fully state-selected ions \cite{tong10a}.

We acknowledge support from the Swiss National Science Foundation (grant no. PP0022\_118921), the University of Basel and the COST Action MP1001 ``Ion Traps for Tomorrow's Applications''. We thank Dr. Nuria Plattner for helpful discussions.

\bibliographystyle{apsrev4-1}
\bibliography{../../../docs/bib_file/main}

\end{document}